%% file: main.tex
\documentclass[unnumsec,webpdf,contemporary,large]{oup-authoring-template}%

\providecommand{\postcode}[1]{#1}
\providecommand{\city}[1]{#1}
\providecommand{\country}[1]{#1}

\usepackage{graphicx}
\graphicspath{{Fig/}}
\usepackage{xcolor,colortbl}
\usepackage{graphics}
\usepackage{amsmath, bm}
\usepackage{tabularray}
\usepackage{bm} 
\usepackage{hyperref}
\usepackage{amsmath}
\usepackage{amssymb}
\usepackage{pifont}
\usepackage{booktabs}
\usepackage{multirow}
\usepackage{subfigure}
\usepackage{subcaption}
\usepackage{makecell}
\usepackage[normalem]{ulem}

\graphicspath{{Fig/}}

\theoremstyle{thmstyletwo}%
\theoremstyle{thmstylethree}%

\begin{document}

\journaltitle{Journal Title Here}
\DOI{DOI HERE}
\copyrightyear{2022}
\pubyear{2019}
\access{Advance Access Publication Date: Day Month Year}
\appnotes{Paper}

\firstpage{1}

\title[Short Article Title]{SGAC: A Graph Neural Network Framework for Imbalanced and Structure-Aware AMP Classification}

\author[1]{Yingxu Wang}
\author[2]{Victor Liang}
\author[3]{Nan Yin}
\author[2]{Siwei Liu$^\ast$}
\author[1,4]{Eran Segal$^\ast$}

\authormark{Wang et al.}

\address[1]{\orgdiv{Department of Machine Learning}, \orgname{Mohamed bin Zayed University of Artificial Intelligence}, \orgaddress{\street{AI Diyafah St}, \postcode{7909}, \city{Abu Dhabi}, \country{United Arab Emirates}}}
\address[2]{\orgdiv{School of Natural \& Computing Science}, \orgname{University of Aberdeen}, \orgaddress{\street{32 Elphinstone Rd}, \postcode{AB24 3EU}, \state{Scotland}, \country{United Kingdom}}}
\address[3]{\orgdiv{Department of Computer Science and Engineering}, \orgname{Hong Kong University of Science and Technology }, \orgaddress{\state{Hong Kong}, \country{China}}}
\address[4]{\orgdiv{Department of
Molecular Cell Biology}, \orgname{Weizmann Institute of Science}, \orgaddress{\state{Rehovot}, \country{Israel}}}
\corresp[$\ast$]{Corresponding author. \href{email:email-id.com}{siwei.liu@abdn.ac.uk, Eran.Segal@weizmann.ac.il}}

\received{Date}{0}{Year}
\revised{Date}{0}{Year}
\accepted{Date}{0}{Year}
\def\method{SGAC}


\abstract{
Classifying Antimicrobial Peptides (AMPs) from the vast collection of peptides derived from metagenomic sequencing offers a promising avenue for combating antibiotic resistance. However, most existing AMP classification methods rely primarily on sequence-based representations and fail to capture the spatial structural information critical for accurate identification. Although recent graph-based approaches attempt to incorporate structural information, they typically construct residue- or atom-level graphs that introduce redundant atomic details and increase structural complexity. Furthermore, the class imbalance between the small number of known AMPs and the abundant non-AMPs significantly hinders predictive performance. To address these challenges, we employ lightweight OmegaFold to predict the three-dimensional structures of peptides and construct peptide graphs using C$_\alpha$ atoms to capture their backbone geometry and spatial topology. Building on this representation, we propose the \textbf{S}patial \textbf{G}NN-based \textbf{A}MP \textbf{C}lassifier (\method{}), a novel framework that leverages Graph Neural Networks (GNNs) to extract structural features and generate discriminative graph representations. To handle class imbalance, \method{} incorporates Weight-enhanced Contrastive Learning to cluster structurally similar peptides and separate dissimilar ones through adaptive weighting, and applies Weight-enhanced Pseudo-label Distillation to generate high-confidence pseudo labels for unlabeled samples, achieving balanced and consistent representation learning. Experiments on publicly available AMP and non-AMP datasets demonstrate that \method{} significantly achieves state-of-the-art performance compared to baselines.}



\maketitle
\input{1_introduction}
\input{3_method}

\input{4_experiment}

\input{5_conclusion}

\bibliographystyle{plain}
\bibliography{reference}

\newpage
\onecolumn

\end{document}

%% file: 1_introduction.tex
\section{1. Introduction}

The increasing prevalence of antibiotic resistance~\cite{salam2023antimicrobial,walsh2023antimicrobial} has intensified the demand for new antimicrobial drugs. Antimicrobial peptides (AMPs), which are short peptides with broad-spectrum antimicrobial activity, are predominantly 10 to 50 amino acids in length as reported in public databases. AMPs are regarded as successors that can “surpass their predecessors" due to their unique antimicrobial mechanisms and broad-spectrum activity~\cite{magana2020value,vishnepolsky2022comparative,yan2020deep,bucataru2024antimicrobial}. In recent years, advanced computational methods such as artificial intelligence and molecular simulation have been extensively applied in the field of AMP development, offering new opportunities for the research and development of these novel antimicrobial molecules~\cite{ma2022identification,huang2023identification,torres2024mining}.

During the evolutionary process of microorganisms, they engage in mutual competition, leading to the production of numerous substances that can resist other microorganisms. The genetic fragments of microorganisms contain many potential sequences that could be encoded as antimicrobial peptides, forming a rich reservoir of antimicrobial peptide sources~\cite{bosch2021antimicrobial,suneja2019microbiome}. Numerous studies have utilized metagenomic sequencing to mine potential antimicrobial peptides~\cite{yang2024review,kumar2024review,cui2024mining}: beginning with the collection of metagenomic sequencing data from diverse sources to identify open reading frames (ORFs)~\cite{couso2017classification} within these sequences—those that have the potential to encode proteins. Following this, classifiers are trained using historical data of AMPs and non-AMPs to learn the patterns characteristic of AMPs. Subsequently, these well-trained classifiers are employed to sort through the open reading frames and categorize the potential AMPs. For example, Ma et al.~\cite{ma2022identification} integrated multiple models, including LSTM~\cite{yu2019review} and BERT~\cite{devlin2018bert}, to construct an AMP mining pipeline. From 154,723 metagenomic sequencing data, they identified 2,349 potential AMPs and synthesized 216 of them, with 181 exhibiting antimicrobial activity. Similarly, Torres et al.~\cite{torres2024mining} used MetaProdigal~\cite{hyatt2012gene}, CD-Hit~\cite{fu2012cd}, and SmORFinder~\cite{durrant2021automated} to screen 1,773 human gut metagenomes, identifying 444,054 potential peptides, and predicted their antimicrobial activity using models such as random forests~\cite{bhadra2018ampep} and BERT, ultimately determining 323 candidate AMPs. They selected 78 of these for synthesis and in vitro experiments, which confirmed that 55 exhibited antimicrobial activity. Compared to the previous two studies, Santos-Jnior et al.~\cite{santos2024discovery} utilized a larger dataset, leveraging 63,410 public metagenomes and 87,920 high-quality prokaryotic genomes. Initially, they employed the MetaProdigal~\cite{hyatt2012gene} tool to predict all ORFs within the metagenomes. Following this, they used the CD-Hit~\cite{fu2012cd} tool to cluster the predicted smORFs, identifying non-redundant peptide sequence families. Finally, they applied the Macrel tool~\cite{santos2020macrel}, a random forest-based machine learning pipeline, to predict potential antimicrobial peptides from large peptide datasets, resulting in 863,498 non-redundant candidate antimicrobial peptides. They selected 100 of these for synthesis and in vitro experiments, which confirmed that 79 exhibited antimicrobial activity.


{Previous research has shown that identifying effective AMPs is a complex process that involves both computational screening and resource-intensive in vitro validation \cite{liu2022efficient,xu2024based,arakal2023silico}. To reduce this experimental burden, earlier studies have primarily relied on sequence-based models such as LSTM~\cite{ma2022identification} and Random Forest~\cite{santos2024discovery} to discover distinguish AMPs and non-AMPs. Although these methods achieve moderate success, they fail to capture the essential spatial organization and geometric dependencies underlying peptide bioactivity. Recent advances in graph-based learning methods typically utilize pretrained protein language models (PLMs) such as AlphaFold2~\cite{jumper2021highly} and ESMFold~\cite{vishnepolsky2022comparative} to predict the 3D spatial structures of peptides and then construct graphs based on residue- or all-atom-level representations~\cite{martinez2024examining, fernandez2024autopeptideml, sun2025multimodal}. However, when applied to large-scale peptide datasets, this pipeline incurs extremely high preprocessing costs for structure prediction and graph construction~\cite{hao2024pgat,chen2024tp}. In addition, the constructed graphs are typically dense and contain redundant atomic details, which increase structural complexity and substantially raise the computational burden of downstream graph representation learning~\cite{yan2023samppred,yim2023se,fernandes2023geometric}. Beyond the architectural limitations of existing models, a fundamental challenge in AMP prediction arises from the severe class imbalance inherent in available datasets, where AMPs constitute only a small fraction of the overall peptide population~\cite{chen2024amppred,chen2025uniamp}. This imbalance tends to bias the learning process toward the dominant non-AMP class, limiting the model’s ability to generalize to rare yet biologically important samples. As a result, existing methods often exhibit reduced discriminative power and unstable performance when applied to large-scale or real-world peptide applications.}

{To address the limitations of existing methods, we first predict the 3D spatial structure of each peptide using a lightweight pretrained protein language model, OmegaFold~\cite{wu2022high}. From the predicted structures, we construct peptide graphs using C$_\alpha$ atoms to capture backbone geometry and spatial topology for accurate AMP classification. Then, we propose a novel framework, \method{}, which employs a Graph Neural Network (GNN) as the backbone encoder to learn spatial characteristics and generate discriminative graph representations for each peptide graph. To enhance representation learning under class imbalance, \method{} introduces two complementary modules: Weight-enhanced Contrastive Learning and Weight-enhanced Pseudo-label Learning. The contrastive module assigns adaptive weights to peptide pairs, improving the model’s ability to distinguish positive and negative samples by clustering structurally similar peptides and separating dissimilar ones, thereby strengthening AMP-specific structural discrimination. The pseudo-label module incorporates weight-aware optimization to generate high-confidence pseudo labels for unlabeled peptides, ensuring semantic consistency between labeled and unlabeled data while refining the representation of minority AMP samples.}

The contributions of our paper can be summarized as follows:
\begin{itemize}
\item {We propose a lightweight peptide graph construction mechanism for spatial AMP prediction, which predicts 3D structures using OmegaFold and constructs graphs based on C$_\alpha$ atoms, effectively capturing backbone geometry while minimizing redundant atomic details.}
\item We develop \method{}, which employs GNNs to encode spatial characteristics of each peptide and integrates Weight-enhanced Contrastive Learning and Pseudo-label Distillation to address the class imbalance and improve classification accuracy.
\item Experiments on publicly available AMP and non-AMP datasets demonstrate that our \method{} significantly outperforms traditional sequence-based methods and achieves state-of-the-art performance, validating its effectiveness in AMP classification.
\end{itemize}

%% file: 3_method.tex
\section{2. Methodology}

\begin{figure*}
    \centering
    \includegraphics[width=0.9\linewidth]{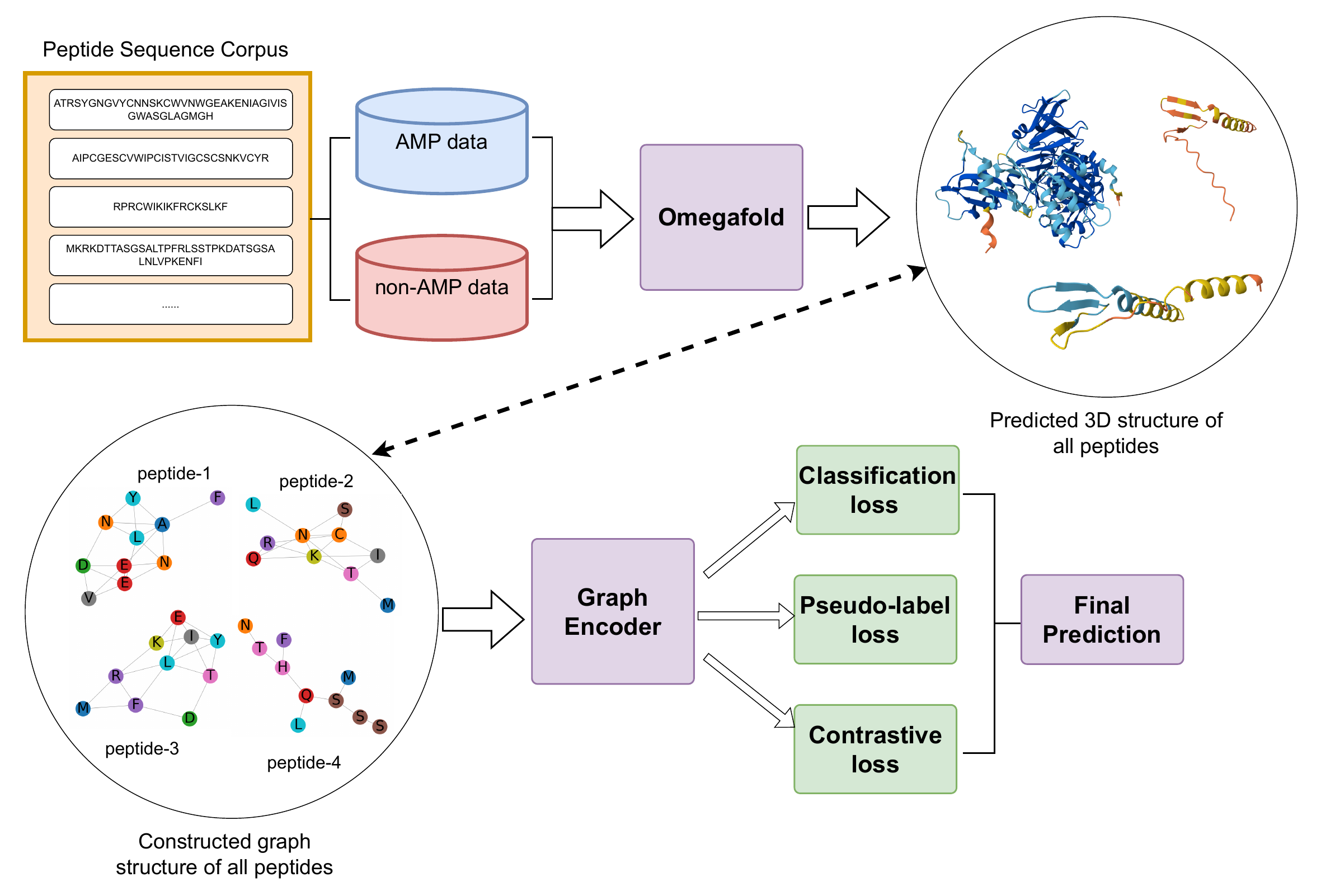}
    \caption{Overall framework of our \method{} model. The framework consists of multiple stages designed to enhance AMP classification. Initially, Omegafold is employed to predict the three-dimensional (3D) structure of peptides based on their amino acid sequences, generating peptide graphs where nodes represent C$_\alpha$ atoms. These graphs are then processed by a Graph Neural Network (GNN) encoder to capture structural and relational features. Then, the embeddings produced by Graph Encoder are refined using three key components: Weight-enhanced Contrastive Learning, which improves feature separation between AMPs and non-AMPs; Weight-enhanced Pseudo-label Distillation, which dynamically refines predictions using high-confidence pseudo labels; and a Classification Loss, which ensures accurate supervised learning by directly optimizing the classification performance. Finally, our \method{} model produce the final prediction results.}
    \label{fig:framework}

\end{figure*}
In this paper, we focus on improving the prediction accuracy of AMPs and non-AMPs by leveraging the spatial structural information of peptides. We first predict the 3D structures of peptides based on their amino acid sequences using Omegafold \cite{wu2022high} and construct peptide graphs from the C$_\alpha$ positions of the residues \cite{yim2023se,liu2024dynamic,watson2022broadly}. Then, we introduce a novel approach, \method{}, which leverages a Graph Neural Network (GNN) as a \textbf{Graph Encoder} to generate comprehensive peptide representations. To address the class imbalance issue and refine predictions, \textbf{Weight-enhanced Contrastive Learning} clusters similar peptides and separates dissimilar ones by assigning specific weights to positive and negative pairs. Furthermore, \textbf{Weight-enhanced Pseudo-label Distillation} dynamically generates reliable and high confidence pseudo labels, ensuring balanced representation of AMPs and non-AMPs while enforcing consistency. The overview of the above process is shown in Figure \ref{fig:framework}.

\subsection{2.1 Graph Construction}\label{sec:3.1}

Let $P = \{a_1, a_2, \dots, a_n\}$ represents a peptide sequence consisting of $n$ amino acids. The AMP and non-AMP data are initially provided as primary amino acid sequences without atomic-level structural details. To address this limitation, we utilize Omegafold \cite{wu2022high}, a pre-trained protein language model that can infer the three-dimensional spatial coordinates of each residue from its amino acid sequence. By leveraging both AMP and non-AMP data by Omegafold, we obtain the corresponding 3D spatial structures for each amino acid residue $a_i$. To focus on structural analysis, we only extract the positions of the C$_\alpha$ atoms, which serve as representative spatial coordinates for the amino acid backbone. Such a process will generate a set of 3D coordinates denoted as $ \mathbf{R} =\{ \mathbf{r}_1, \mathbf{r}_2, \dots, \mathbf{r}_n \} \in \mathbb{R} ^{N \times 3}$, where $\mathbf{r}_i \in \mathbb{R}^3$ corresponds to the spatial position of the C$_\alpha$ atom for amino acid $a_i$.

To represent the peptides as graphs, we use an undirected graph $G = (V, E, \mathbf{X})$ to represent each peptide, where $V$ denotes the set of nodes, $E$ the set of edges, and $\mathbf{X}$ the node feature matrix. The nodes $V$ correspond to the C$_\alpha$ atoms of the amino acids, defined as $V = \{v_1, v_2, \dots, v_n\}$, with each node $v_i$ representing the C$_\alpha$ atom of amino acid $a_i$. Edges in the graph are determined based on the Euclidean distances between C$_\alpha$ atoms. Specifically, an edge is established between nodes $v_i$ and $v_j$ if the distance between their corresponding C$_\alpha$ atoms is less than a predefined threshold $\delta$:

\begin{equation}
E=\left\{\left(v_i, v_j\right) \mid\left\|\mathbf{r}_i-\mathbf{r}_j\right\|_2<\delta, i \neq j\right\},  
\end{equation}
where $\delta$ controls the connectivity of the graph. The node feature matrix $\mathbf{X}$ encodes the biochemical identity of each amino acid. We produce a comprehensive set of unique amino acid types from the AMP and non-AMP datasets to generate these features. Each amino acid is then represented as a one-hot encoded vector based on its sequence identity, ensuring that the feature matrix $\mathbf{X}$ captures essential sequence-based information.

The constructed graph $G$ captures the spatial relationships between amino acids within the peptide, providing a robust framework for analysing protein structure and function. Additionally, the focus on spatial connectivity through the C$_\alpha$ atoms ensures that the graph $G$ captures essential structural features of each peptide, which are critical for distinguishing AMPs from non-AMPs.

\subsection{2.2 Graph Encoder}\label{sec:3.2}

Peptides are intricate biomolecules whose functions are deeply influenced by their structure and spatial relationships. To accurately distinguish AMPs from non-AMPs, we transform the input peptide graphs into graph representations to capture essential structural and relational information within peptides. This approach leverages peptides' intricate spatial patterns and connectivity to enhance predictive accuracy and biological relevance.

Given a peptide graph $G = (V, E, \mathbf{X})$, we leverage a Graph Neural Network (GNN) to encode its structure and node attributes. Let $\mathbf{h}_v^{(l)}$ represent the embedding for node $v$ at the $l$ th layer of the network. The initial embedding $\mathbf{h}_v^{(0)}$ is defined by $\mathbf{h}_v^{(0)} = \text{MLP}(\mathbf{x}_v)$, where the MLP$(\cdot)$ is a learnable multi-layer perceptron that maps the node feature $\mathbf{x}_v$ into the latent space. Subsequently, for each node $v \in V$, the embedding $\mathbf{h}_v^{(l)}$ is updated iteratively through a message-passing framework. At each iteration, the embedding of $v$ is updated by aggregating information from its neighboring nodes $\mathcal{N}(v)$ from the previous layer, and this aggregated information is combined with the embedding of $v$ from the same layer. Formally, this update process is defined as:
\begin{equation}
\mathbf{h}_{v}^{l}= \mathcal{C}^{(l-1)}\left(\mathbf{h}_{v}^{l-1}, \mathcal{A}^{(l-1)} \left(\left\{\mathbf{h}_{u}^{l-1}\right\}_{u \in \mathcal{N}(v)}\right) \right),
\end{equation}
where $\mathcal{A}^{(l-1)}$ denotes the aggregation operation at the $l-1$-th layer, which aggregates embeddings from the neighbors of $v$, and $\mathcal{C}^{(l-1)}$ represents the combination operation at the $l-1$-th layer, which integrates the aggregated neighbor information with the embedding of $v$ itself. This iterative message-passing mechanism enables the GNN to effectively capture both the local structural interactions among amino acids and the global topology of the peptide graph.

After $L$ layers of iterative updates, a graph-level representation $\mathbf{z}$ is computed for each peptide graph $G$ using a readout function, which aggregates the node representations at the final layer:
\begin{equation}
\mathbf{z}= F\left(G\right)=\operatorname{READOUT}\left(\left\{\mathbf{h}_{v}^{L}\right\}_{v \in \mathcal{V}}\right),
\end{equation} 
where $\mathbf{z}$ is the graph-level representation for each peptide generated by the GNN function $F(\cdot)$. The $\operatorname{READOUT}$ function can be implemented in various strategies, such as summation, averaging, maximization of node embeddings \cite{xu2018powerful,wang2024dusego,yao2023improving} or incorporating a virtual node \cite{li2015gated,wang2025protomol,wang2025nested} to capture global graph information. The graph-level representation $\mathbf{z}$ captures the peptide’s structural and relational characteristics, providing a robust foundation for distinguishing AMPs from non-AMPs. 

\subsection{2.3 Weight-enhanced Contrastive Learning}\label{sec:contrastive learning}

Contrastive learning enhances the discriminative power of feature representations by aligning labeled similar samples while ensuring separation from dissimilar ones \cite{chen2023heterogeneous,wang2023cl4ctr}, which is effective for distinguishing AMPs from non-AMPs by capturing their unique structural and relational differences. However, a critical challenge in AMP classification lies in the class imbalance, where non-AMP samples significantly outnumber AMP samples. This imbalance can lead standard contrastive learning methods to overemphasize the majority class (non-AMPs), thereby degrading the model's ability to classify the minority class (AMPs) accurately.

To address this issue, we introduce Weight-enhanced Contrastive Learning, a tailored approach that incorporates class-specific weights to balance the influence of positive and negative samples during training. By appropriately weighting contributions from each class, this method mitigates the effects of class imbalance, ensuring that the learned representations reflect the distinguishing features of AMPs while maintaining robust discrimination against non-AMPs. 

Assuming the Graph Encoder has $L$ message passing layers, we first extract the graph embeddings of the $L$-th layer, where $\mathbf{Z}^L = [\mathbf{z}_1^L, \mathbf{z}_2^L, \dots, \mathbf{z}_{|\mathcal{N}_\text{label}|}^L]^T \in \mathbb{R}^{{|\mathcal{N}_\text{label}|} \times d}$ for $|\mathcal{N}_\text{label}|$ labeled peptide graphs, where $\mathbf{z}_i^L \in \mathbb{R}^d$ is the $d$-dimensional embedding of the $i$-th labeled peptide, and the corresponding label vector $\mathbf{y} = [y_1, y_2, \dots, y_{|\mathcal{N}_\text{label}|}]$. Contrastive learning aims to optimize embeddings such that peptides sharing the same label are positioned closer together in the latent space. In contrast, those with different labels are pushed further apart. To achieve this, the pairwise Euclidean distance between embeddings is computed as:

\begin{equation}
D_{i j}=\left\|\mathbf{z}_i^L-\mathbf{z}_j^L\right\|_2, \quad \forall i, j \in\{1,2, \ldots, {|\mathcal{N}_\text{label}|}\}.
\end{equation}

To address the issue of class imbalance, we incorporate class-specific weights, denoted as $\omega_{cl}^{\text{pos}}$ for positive pairs (samples with the same label) and $\omega_{cl}^{\text{neg}}$ for negative pairs (samples with different labels). Specifically, $\omega_{cl}^{\text {pos }}$ is calculated as:

\begin{equation}
\label{eq:5}
\omega_{cl}^{\text{pos}} = 1-\frac{\left|\mathcal{N}_\text{label}^{\text {pos }}\right|}{\left|\mathcal{N}_\text{label}^{\text {pos }}\right|+\left|\mathcal{N}_\text{label}^{\text{neg}}\right|},
\end{equation}
while $\omega_{cl}^{\text {neg }}$ is calculated by:
\begin{equation}
\label{eq:6}
\omega_{cl}^{\text{neg}} = 1-\frac{\left|\mathcal{N}_\text{label}^{\text {neg }}\right|}{\left|\mathcal{N}_\text{label}^{\text {pos }}\right|+\left|\mathcal{N}_\text{label}^{\text {neg }}\right|},
\end{equation}
where $\mathcal{N}_\text{label} = \mathcal{N}_\text{label}^{\text {pos }} + \mathcal{N}_\text{label}^{\text {neg}}$, $\mathcal{N}_\text{label}^{\text {pos }}$ and $\mathcal{N}_\text{label}^{\text {neg }}$ denote the sets of AMP and non-AMP samples in the labeled dataset, respectively, and $\left|\mathcal{N}_\text{label}^{\text {pos }}\right|$ and $\left|\mathcal{N}_\text{label}^{\text {neg }}\right|$ represent their respective sizes. 

These weights ensure that both classes contribute proportionally to the learning process. The positive contrastive loss, which encourages embeddings of peptides with the same label to cluster together, is defined as:
\begin{equation}
\begin{aligned} \mathcal{L}_{c l}^{\mathrm{pos}}=\omega_{c l}^{\mathrm{pos}} \sum_{i=1}^{|\mathcal{N}_\text{label}|}  \sum_{j=1}^{|\mathcal{N}_\text{label}|} &\mathbb{I}\left(y_i=y_j\right) D_{i j}^2, \\ & \forall i, j \in\left\{1,2, \ldots,\left|\mathcal{N}_\text{label}\right|\right\},
\end{aligned}
\end{equation}
where $\mathbb{I}(\cdot)$ is an indicator function that equals 1 if $y_i = y_j$, and 0 otherwise. This formulation allows for the effective integration of class-specific contributions, improving the model's ability to capture the intrinsic relationships within the peptide data, even when the class distribution is imbalanced.

To ensure sufficient separation between embeddings of peptides with different labels, we apply a margin $m$ for negative pairs. This margin enforces a minimum distance between dissimilar samples, thereby enhancing the discriminative power of the model. The negative contrastive loss is defined as:
\begin{equation}
\begin{aligned} \mathcal{L}_{c l}^{\mathrm{neg}}=\omega_{c l}^{\mathrm{neg}} \sum_{i=1}^{|\mathcal{N}_\text{label}|} \sum_{j=1}^{|\mathcal{N}_\text{label}|}  \mathbb{I}&\left(y_i \neq y_j\right) \max \left(0, m-D_{i j}\right)^2, \\ & \forall i, j \in\left\{1,2, \ldots,\left|\mathcal{N}_\text{label}\right|\right\},\end{aligned}
\end{equation}
where $\mathbb{I}\left(y_i \neq y_j\right)$ is an indicator function that equals 1 if $y_i \neq y_j$ and 0 otherwise. The term $\max \left(0, m-D_{i j}\right)$ ensures that negative pairs contribute to the loss only when their distance $D_{i j}$ is less than the margin $m$.

Finally, the overall Weight-enhanced Contrastive Loss, $\mathcal{L}_{\text{cl}}$, combines the positive and negative losses, normalized by the total number of sample pairs to account for the dataset size, which is defined as follows:
\begin{equation}
\mathcal{L}_{\text {cl}}=\frac{1}{|\mathcal{N}_\text{label}|*(|\mathcal{N}_\text{label}|-1)}\left(\mathcal{L}_{cl}^{\mathrm{pos}}+\mathcal{L}_{cl}^{\mathrm{neg}}\right).    
\end{equation}
This formulation ensures a balanced optimization of embeddings, effectively leveraging both positive and negative pairs while mitigating the effects of class imbalance.

\subsection{2.4 Weight-enhanced Pseudo Label Distillation}

Pseudo-label learning leverages unlabeled data by assigning high-confidence pseudo-labels to enable iterative refinement of the model's understanding of the data distribution~\cite{wang2024degree}. This process ensures semantic consistency across different latent spaces and is particularly useful for addressing the scarcity of labeled AMP data while facilitating the discovery of novel AMP sequences. However, traditional pseudo-label distillation methods often suffer from the inherent class imbalance between AMP and non-AMP samples, resulting in model bias toward the majority class.

To address this issue, we propose Weight-Enhanced Pseudo-Label Distillation, a method designed to dynamically balance the contributions of AMP and non-AMP samples while aligning semantic information between labeled and unlabeled data. Additionally, this method incorporates soft-label refinement to produce accurate peptide representations, effectively mitigating the impact of class imbalance.

Assuming the Graph Encoder consists of $L$ message-passing layers, we produce pseudo-labels for unlabeled peptides using the embeddings generated from the $l$-th hidden layer. These embeddings are represented as $\mathbf{Z}^l=\left[\mathbf{z}_1^l, \mathbf{z}_2^l, \ldots,\left.\mathbf{z}_{\left|\mathcal{N}_{\text {unlabel }}\right|}^l\right|^T \in \mathbb{R}^{\left|\mathcal{N}_{\text {unlabel}}\right| \times d}\right.$, where $\left|\mathcal{N}_{\text {unlabel}}\right|$ denotes the number of unlabeled peptide graphs, and $\mathbf{z}_i^l \in \mathbb{R}^d$ is the $d$-dimensional embedding of the $i$-th unlabeled peptide.

We cluster these peptide embeddings into $K$ clusters\footnote{ $K=2$, representing AMP and non-AMP classes} using the K-Means algorithm \cite{hartigan1979k}. The initial cluster centers are denoted as $\mathbf{C}=\left\{\mathbf{c}_1, \mathbf{c}_2, \ldots, \mathbf{c}_K\right\} \in \mathbb{R}^{K \times d}$. The pairwise distance $V_{i j}$ between the $i$-th peptide embedding $\mathbf{z}_i^l$ and the $j$-th cluster center $\mathbf{c}_j$ is computed as:
\begin{equation}
\begin{aligned} V_{i j}&=\left\|\mathbf{z}_i^l-\mathbf{c}_j\right\|_2, \\ & \quad \forall i \in\left\{1,2, \ldots,\left|\mathcal{N}_{\text {unlabel}}\right|\right\}, \\ & \quad \forall j \in\{1,2, \ldots, K\},
\end{aligned}
\end{equation}
where $V_{i j}$ denotes the Euclidean distance between $\mathbf{z}_i^l$ and $\mathbf{c}_j$. Then, the distances are then converted into probabilities using a softmax function:
\begin{equation}
p_{i j}=\frac{\exp \left(-V_{i j}\right)}{\sum_{n=1}^K \exp \left(-V_{i n}\right)}.
\end{equation}
Finally, the pseudo-label $\hat{y}_i$ for $i$-th unlabeled peptide is assigned to the class with the highest probability: 
\begin{equation}
\hat{y}_i=\arg \max _j p_{i j}.
\end{equation}

Moreover, to emphasize high-confidence predictions, we define the associated confidence score for each unlabeled peptide. In details, for $i$-th unlabeled peptide, the confidence score $s_i$ is calculated as
\begin{equation}
s_i=\max _j p_{i j},
\end{equation}
which can reflect the model's certainty in its prediction. We then construct the confident sets for the unlabeled peptides based on their pseudo labels and confidence scores as follows:
\begin{equation}
\begin{aligned}
\mathcal{N}_{\mathrm{c}}^{\text {pos }} & =\left\{i \mid \hat{y}_i=1, s_i>\tau\right\}, \\
\mathcal{N}_{\mathrm{c}}^{\text {neg }} & =\left\{j \mid \hat{y}_j=0, s_j>\tau\right\}, \\
& \forall i, j \in\left\{1,2, \ldots,\left|\mathcal{N}_{\text {unlabel}}\right|\right\},
\end{aligned}
\end{equation}
where $\mathcal{N}_{\mathrm{c}}^{\text {pos }}$ represents the confident set of peptides with the pseudo label $\hat{y}=1$, and $\mathcal{N}_{\mathrm{c}}^{\text {neg }}$ represents the confident set of peptides with the pseudo label $\hat{y}=0$. Here, $\tau$ is a predefined threshold value, which is set to 0.5 for convenience. These confident sets form the foundation for further training and refinement in the AMP classification task. 

For further adapting to evolving embedding distributions, the cluster centers are updated based on mean embeddings of the confident samples assigned to each cluster. For cluster $j$, the new cluster center $\mathbf{\hat{c}}_j$ is computed as:

\begin{equation}
\begin{aligned}
\mathbf{\hat{c}}_j=\frac{1}{\left|\mathcal{N}_c^{j}\right|} &\sum_{i \in \mathcal{N}_c^j} \mathbf{z}_i^l,\\
&\forall j \in \{\text{pos}, \text{neg}\},
\end{aligned}
\end{equation}
where $\mathcal{N}^j_c$ is the set of confident samples assigned to cluster $j$, $|\mathcal{N}^j_c|$ is the size of $\mathcal{N}^j_c$.

Next, to mitigate the class imbalance inherent in AMP and non-AMP data, class-specific weights are introduced, which are denoted as $\omega_{cl}^{\text{pos}}$ for AMP data and $\omega_{cl}^{\text{neg}}$ for non-AMP data. Specifically, $\omega_{pl}^{\text {pos }}$ is calculated as:
\begin{equation}
\label{eq:16}
\omega_{pl}^{\text {pos }} = 1-\frac{\left|\mathcal{N}_{\text {c}}^{\text {pos }}\right|}{\left|\mathcal{N}_{\text {c}}^{\text {pos }}\right|+\left|\mathcal{N}_{\text {c}}^{\text {neg }}\right|},
\end{equation}
similarly, $\omega_{pl}^{\text {neg }}$ is calculated by: 
\begin{equation}
\label{eq:17}
\omega_{pl}^{\text{neg}} = 1-\frac{\left|\mathcal{N}_{\text{c}}^{\text {neg }}\right|}{\left|\mathcal{N}_{\text{c}}^{\text {pos }}\right|+\left|\mathcal{N}_{\text{c}}^{\text {neg }}\right|}.
\end{equation}

The positive pseudo label  loss, $\mathcal{L}^{\text {pos}}_{pl
}$, is calculated as:

\begin{equation}
\mathcal{L}_{pl}^{\mathrm{pos}}=\omega_{pl}^{\mathrm{pos}} \frac{1}{|\mathcal{N}_{c}^\text{pos}|} \sum_{i \in \mathcal{N}_{c}^\text{pos}}-\log \left(p_{i, \hat y = 1}\right),
\end{equation}
where $p_{i, \hat{y}=1}$ denotes the predicted probability of the $i$-th unlabeled peptides in $\mathcal{N}_{c
}^\text{pos}$. Conversely, the negative pseudo label loss, $\mathcal{L}_{pl}^{\text {neg}}$, is expressed as:

\begin{equation}
\mathcal{L}_{pl
}^{\text {neg }}=\omega_{pl
}^{\text {neg }} \frac{1}{|\mathcal{N}_{c
}^\text{neg}|} \sum_{j \in \mathcal{N}_{c
}^\text{neg}}-\log \left(p_{j, \hat y = 0}\right),
\end{equation}
where $p_{j, \hat{y}=0}$ is the predicted probability of the $j$-th unlabeled peptides in $\mathcal{N}_{c
}^\text{neg}$.

Furthermore, a soft-label refinement loss is introduced using Kullback-Leibler (KL) divergence to refine pseudo-labels and maintain consistency across different layers of Graph Encoder. The soft-label refinement loss is defined as:

\begin{equation}
\mathcal{L}_{\mathrm{soft}}=\alpha \frac{1}{\left|\mathcal{N}_{\mathrm{c}}\right|} \sum_{i=1}^{\left|\mathcal{N}_{\mathrm{c}}\right|} \sum_{j=1}^K {p}_{i j} \log \left(\frac{{p}_{i j}}{\hat{p}_{i j}}\right),
\end{equation}
where $\mathcal{N}_\text{c} = \mathcal{N}_\text{c}^\text{pos} \cup \mathcal{N}_\text{c}^\text{neg}$, $|\mathcal{N}_\text{c}|$ is the size of $\mathcal{N}_\text{c}$, ${p}_{i j}$ is the probability of the $i$-th unlabeled peptides belonging to class $j$ from layer $l$, $\hat{p}_{i j}$ is the predicted probability of the $i$-th unlabeled peptides belonging to class $j$ from layer $L$, and $\alpha=1-\bar{c}$ is a dynamic weighting factor influenced by the mean confidence, which is defined as
\begin{equation}
\bar{c}=\frac{1}{\left|\mathcal{N}_{\mathrm{c}}\right|} \sum_{i=1}^{\left|\mathcal{N}_{\mathrm{c}}\right|} \max _j p_{i j}.    
\end{equation}

The overall Weight-enhanced Pseudo-label loss, $\mathcal{L}_{\mathrm{pl}}$, integrates the above three components:
\begin{equation}
\mathcal{L}_{\text {pl}}=\mathcal{L}_{pl}^{\text {pos }}+\mathcal{L}_{pl}^{\text {neg }}+\mathcal{L}_{\text {soft }}.
\end{equation}

This composite loss ensures balanced pseudo-label learning, aligning the model's predictions with true distributions while addressing class imbalances and refining pseudo-label accuracy.

\subsection{2.5 Learning Framework}

To ensure effective learning from supervised data, we minimize the expected error for labeled data using a classification loss. Assuming the Graph encoder has $L$ message passing layers, we extract the embeddings of the final layer, where $\mathbf{Z}^L = [\mathbf{z}_1^L, \mathbf{z}_2^L, \dots, \mathbf{z}_{|\mathcal{N}_{\text {label}}|}^L]^T \in \mathbb{R}^{{|\mathcal{N}_{\text {label}}|} \times d}$ for $|\mathcal{N}_{\text {label}}|$ labeled peptide graphs, where $\mathbf{z}_i^L \in \mathbb{R}^d$ is the $d$-dimensional embedding of the $i$-th peptides, and the corresponding label vector $\mathbf{y} = [y_1, y_2, \dots, y_{|\mathcal{N}_{\text {label}}|}]$. The classification loss is defined as:
\begin{equation} 
\mathcal{L}_{\text{cf}} = \frac{1}{|\mathcal{N}_{\text {label}}|} \sum_{i=1}^{|\mathcal{N}_{\text {label}}|} \text{CE}(\operatorname{MLP}(\mathbf{z}^L_i),y_i), 
\end{equation}
where MLP $(\cdot)$ is a multi-layer perceptron that maps the protein embedding $\mathbf{z}^L_i$ to the predicted label space, and $\text{CE}(\cdot, \cdot)$ represents the cross-entropy loss function.

Finally, the overall training objective of \method{} combines the classification loss $\mathcal{L}_{\mathrm{cf}}$, the Weight-enhanced Contrastive Learning loss $\mathcal{L}_{\mathrm{cl}}$ and the Weight-enhanced Pseudo-label loss $\mathcal{L}_{\text {pl }}$, which is formulated as follows:

\begin{equation} 
\mathcal{L} = \mathcal{L}_{\text{cf}} + \lambda \mathcal{L}_{\text{cl}} + \gamma  \mathcal{L}_{\text{pl}}, \label{eq:loss} 
\end{equation}
where $\lambda$ and $\gamma$ are hyperparameters that balance the ratio of the Weight-enhanced Contrastive Learning loss and the Weight-enhanced Pseudo-label loss relative to the classification loss. These components enable our \method{} to effectively integrate labeled data, structural information, and pseudo-labeled samples, ensuring robust and balanced learning for AMP and non-AMP classification.

\subsection{2.6 AMP Prediction}
\label{sec:amp_prediction}

In the final stage of our framework, we predict whether a given peptide sequence belongs to the AMP class or the non-AMP class based on the learned graph representations. The AMP prediction process leverages the comprehensive features captured by the Graph Encoder, as well as the refinements provided by Weight-enhanced Contrastive Learning and Weight-enhanced Pseudo-label Distillation.

Given the graph-level representation $\mathbf{z}_i$ of $i$-th peptide graph generated by the Graph Encoder, we employ a Multi-Layer Perceptron (MLP) for classification. Specifically, the prediction probability $p_{ij}$ is computed as:

\begin{equation}
p_{ij} = \mathrm{softmax}(\mathrm{MLP}(\mathbf{z}_i)),
\end{equation}
where the MLP $(\cdot)$ maps the graph representation $\mathbf{z}_i$ to a probability distribution over the AMP and non-AMP classes.

The final classification decision for $i$-th peptide is determined by:

\begin{equation}
\overline{y}_i = \arg \max _j p_{i j} ,
\end{equation}
where $\overline{y}_i$ is the predicted label for the $i$-th peptide.

%% file: 4_experiment.tex
\section{3. Experiments}

In this section, we conduct extensive experiments to verify the effectiveness of \method{} in AMP and non-AMP classification task.

\subsection{3.1 Experimental settings}

\textbf{Dataset.} The AMP data was sourced from the DRAMP database \cite{ma2024dramp}, which provides detailed annotations of antimicrobial peptides, while the non-AMP data was obtained from \cite{ma2022identification}. To ensure consistency, both AMP and non-AMP sequences were filtered to include only peptides with lengths ranging from 10 to 50 amino acids \cite{jha2022prediction, wang2022machine}. Following the graph construction methodology described in Section \ref{sec:3.1}, these amino acid sequences are transformed into graphs using their predicted 3D structures derived from amino acid sequences. The statistics of the processed datasets are presented in Table \ref{tab:statistics}. Additionally, we adopt a standard random stratified split of the dataset into training (70\%), testing (10\%), and validation (20\%) sets.

\begin{table}[t]
    \centering
    \caption{Statistics of the AMP and non-AMP datasets, which presents the number of graphs, average nodes, and average edges for the complete dataset, as well as for AMPs and non-AMPs separately.}
    \resizebox{0.48\textwidth}{!}{%
    \begin{tabular}{lcccc}
        \toprule
        Datasets       & Graphs  & Avg. Nodes & Avg. Edges & Classes \\
        \midrule
        AMPs \& non-AMPs & 65,971  & 38.8       & 195.8      & 2       \\
        AMPs            & 7,697   & 23.1       & 118.6      & 2       \\
        non-AMPs        & 58,274  & 40.9       & 206.1      & 2       \\
        \bottomrule
    \end{tabular}
    }
    \label{tab:statistics}
\end{table}

\textbf{Baselines.} To validate the effectiveness of \method{}, we compare three categories of methods: (1) five sequence-based methods, including BERT ~\citep{devlin2018bert}, Attention ~\citep{vaswani2017attention}, LSTM~\citep{hochreiter1997long}, Random Forest~\citep{breiman2001random}, and Diff-AMP ~\cite{wang2024diff}; (2) three general graph neural networks (GNNs), namely GCN~\citep{kipf2016semi}, GIN~\citep{xu2018powerful} and GraphSAGE~\citep{hamilton2017inductive}; and three recent GNNs-based AMP prediction methods, namely sAMPpred~\cite{yan2023samppred}, MP-BERT~\cite{badrinarayanan2024multi} and PGAT~\cite{hao2024pgat}.

\textbf{Evaluation metrics.} {To comprehensively evaluate the performance of \method{}, we employ two widely used metrics for imbalanced classification tasks, F1 score and Matthews Correlation Coefficient (MCC)~\cite{chicco2023matthews,diallo2024machine}, which provide complementary insights into model performance under class imbalance.}  

\begin{itemize}
    \item \textbf{F1 score} is the harmonic mean of precision and recall, which evaluates the balance between false positives and false negatives. It is particularly effective in imbalanced datasets and is computed as follows:

\begin{equation} 
\mathrm{Precision} = \frac{\mathrm{True\ Positive}}{\mathrm{True\ Positive} + \mathrm{False\ Positive}}, 
\end{equation}

\begin{equation} 
\mathrm{Recall} = \frac{\mathrm{True\ Positive}}{\mathrm{True\ Positive} + \mathrm{False\ Negative}}, 
\end{equation}

\begin{equation} 
\mathrm{F1\ Score} = \frac{2 \times \mathrm{Precision} \times \mathrm{Recall}}{\mathrm{Precision} + \mathrm{Recall}}. 
\end{equation}

A higher F1 score indicates better overall model performance.

\item {\textbf{Matthews Correlation Coefficient (MCC)} is a robust metric that considers all four outcomes of the confusion matrix: true positives (TP), true negatives (TN), false positives (FP), and false negatives (FN). This balanced formulation makes it particularly suitable for evaluating models on imbalanced datasets. It is defined as}:

\begin{equation}
\mathrm{MCC} = \frac{ (TP \times TN) - (FP \times FN) }{\sqrt{(TP + FP)(TP + FN)(TN + FP)(TN + FN)}}.
\end{equation}

{MCC ranges from $-1$ to $+1$, where $+1$ indicates perfect prediction, $0$ corresponds to random guessing, and $-1$ indicates total disagreement. Higher MCC values reflect more reliable performance under imbalance.}

\end{itemize}

\textbf{Implementation details.} {Random Forest is implemented using scikit-learn\footnote{https://scikit-learn.org/}, while all other models, including \method{} and other baselines are implemented using PyTorch\footnote{https://pytorch.org/}. For \method{}, we use GIN \citep{xu2018powerful} as the backbone for the Graph Encoder introduced in Section \ref{sec:3.2}, incorporating a mean-pooling layer as the readout function. All experiments for \method{} and baselines are conducted on NVIDIA A100 GPUs to ensure a fair comparison, where the learning rate of Adam optimizer is set to ${10^{-4}}$, hidden embedding dimension 256, weight decay ${10^{-12}}$, and GNN layers 4. For all methods, we re-trained the models from scratch using the same data splitting strategy, and the performance of each method was evaluated on all samples and reported as the average over five independent runs.}



        

\begin{table*}[t]
\centering
\small
\caption{{Performance comparison among baselines and \method{} across different AMP length intervals and the full dataset. \textbf{Bold} results indicate the best performance.}}
\resizebox{\textwidth}{!}{
\begin{tabular}{clcc|cc|cc|cc}
\toprule
\multirow{2}{*}{\raisebox{-0.3\totalheight}{\textbf{Type}}} 
& \multirow{2}{*}{\raisebox{-0.3\totalheight}{\textbf{Methods}}} 
& \multicolumn{2}{c}{\textbf{10--20}} 
& \multicolumn{2}{c}{\textbf{20--40}} 
& \multicolumn{2}{c}{\textbf{40--50}} 
& \multicolumn{2}{c}{\textbf{Full}} \\
& & \textbf{F1} & \textbf{MCC} & \textbf{F1} & \textbf{MCC} & \textbf{F1} & \textbf{MCC} & \textbf{F1} & \textbf{MCC} \\
\midrule
\multirow{5}{*}{\rotatebox{90}{\makecell[c]{Sequential}}} 
& BERT & 73.43 $\pm$ 0.22 & 56.91 $\pm$ 0.12 & 72.54 $\pm$ 0.37 & 55.58 $\pm$ 0.28 & 70.41 $\pm$ 0.42 & 52.12 $\pm$ 0.18 & 71.33 $\pm$ 0.48 & 54.44 $\pm$ 0.36 \\
& Diff-AMP & 61.95 $\pm$ 1.02 & 38.31 $\pm$ 0.79 & 60.22 $\pm$ 0.76 & 34.17 $\pm$ 1.33 & 58.94 $\pm$ 0.65 & 33.30 $\pm$ 0.56 & 61.08 $\pm$ 0.75 & 37.33 $\pm$ 0.89 \\
& LSTM & 84.76 $\pm$ 0.92 & 69.85 $\pm$ 0.88 & 82.60 $\pm$ 1.06 & 65.30 $\pm$ 0.72 & 79.96 $\pm$ 1.32 & 60.01 $\pm$ 0.67 & 82.09 $\pm$ 0.90 & 66.44 $\pm$ 0.75 \\
& Attention & 83.09 $\pm$ 0.77 & 66.27 $\pm$ 0.65 & 80.61 $\pm$ 1.21 & 63.42 $\pm$ 0.85 & 77.17 $\pm$ 1.09 & 59.09 $\pm$ 0.78 & 81.11 $\pm$ 0.89 & 65.19 $\pm$ 0.71 \\
& Random Forest & 92.21 & 85.03 & 90.64 & 82.45 & 87.01 & 76.03 & 87.97 & 78.08 \\
\midrule
\multirow{3}{*}{\rotatebox{90}{\makecell[c]{General\\Graph}}} 
& GCN & 92.82 $\pm$ 0.55 & 85.08 $\pm$ 0.32 & 84.21 $\pm$ 0.83 & 73.08 $\pm$ 0.48 & 78.98 $\pm$ 0.94 & 61.03 $\pm$ 0.61 & 88.19 $\pm$ 0.68 & 81.22 $\pm$ 0.56 \\
& GraphSAGE & 93.01 $\pm$ 0.76 & 86.11 $\pm$ 0.42 & 85.33 $\pm$ 0.73 & 75.12 $\pm$ 0.55 & 80.17 $\pm$ 0.67 & 63.38 $\pm$ 0.54 & 91.15 $\pm$ 0.71 & 84.08 $\pm$ 0.48 \\
& GIN & 93.72 $\pm$ 0.64 & 86.39 $\pm$ 0.53 & 86.77 $\pm$ 0.49 & 76.39 $\pm$ 0.39 & 83.30 $\pm$ 0.82 & 65.18 $\pm$ 0.76 & 92.12 $\pm$ 0.74 & 85.33 $\pm$ 0.59 \\
\midrule
\multirow{4}{*}{\rotatebox{90}{\makecell[c]{Graph\\based\\AMP}}} 
& sAMPpred & 93.03 $\pm$ 0.55 & 86.55 $\pm$ 0.41 & 90.78 $\pm$ 0.69 & 83.46 $\pm$ 0.48 & 87.12 $\pm$ 0.98 & 79.44 $\pm$ 0.80 & 91.01 $\pm$ 0.73 & 82.39 $\pm$ 0.57 \\
& MP-BERT & 92.66 $\pm$ 0.66 & 87.09 $\pm$ 0.50 & 91.00 $\pm$ 0.43 & 84.88 $\pm$ 0.61 & 89.36 $\pm$ 0.83 & 82.98 $\pm$ 0.73 & 91.53 $\pm$ 0.55 & 85.19 $\pm$ 0.63 \\
& PGAT & 93.91 $\pm$ 0.43 & 87.33 $\pm$ 0.32 & 91.33 $\pm$ 0.56 & 84.67 $\pm$ 0.44 & 89.01 $\pm$ 0.69 & 81.72 $\pm$ 0.52 & 93.04 $\pm$ 0.49 & 86.98 $\pm$ 0.48 \\
& \method{} & \textbf{95.31 $\pm$ 0.59} & \textbf{94.68 $\pm$ 0.48} & \textbf{93.63 $\pm$ 0.70} & \textbf{92.52 $\pm$ 0.52} & \textbf{91.25 $\pm$ 0.81} & \textbf{90.12 $\pm$ 0.49} & \textbf{94.83 $\pm$ 0.77} & \textbf{93.02 $\pm$ 0.56} \\
\bottomrule
\end{tabular}
}
\label{tab:experimental_results}
\end{table*}

\begin{figure*}[t]
    \centering
    \includegraphics[width=1.0\textwidth]{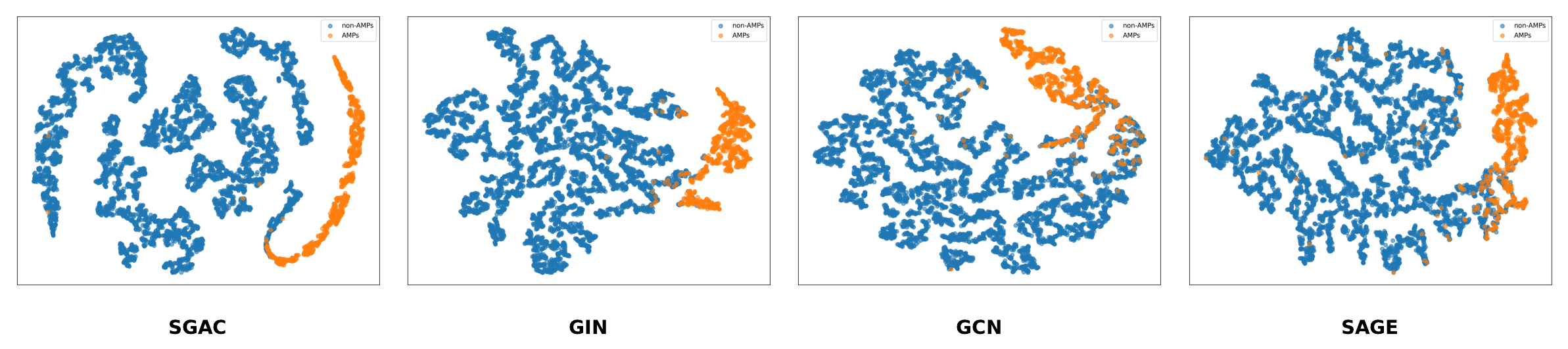} 
    \caption{AMP and non-AMP classification visualization using different models.}
    \label{fig:amp_nonamp_tsne}

\end{figure*}

\begin{table}[t]
    \centering
    \caption{{The performance of \method{} with different GNN
architectures on the full dataset. \textbf{Bold} results indicate the best performance.}}
    \begin{tabular}{lcc}
        \hline
        \textbf{Method} & \textbf{F1} & \textbf{MCC} \\
        \hline
        \method{} w GCN & 92.45 {\scriptsize $\pm$ 0.47} & 91.31 {\scriptsize $\pm$ 0.39} \\
        \method{} w GraphSAGE & 93.75 {\scriptsize $\pm$ 0.64} & 92.58 {\scriptsize $\pm$ 0.43} \\
        \method{} w GIN & \textbf{94.83 {\scriptsize $\pm$ 0.77}} & \textbf{93.02 {\scriptsize $\pm$ 0.56}} \\
        \hline
    \end{tabular}
    \label{tab:different gnn}

\end{table}


\subsection{3.2 Performance comparisons}


{We present the results of the proposed \method{} and all baselines in Table \ref{tab:experimental_results}. We can observe that (1) Random Forest demonstrates the highest performance among sequence-based methods, indicating its strong capability in extracting informative features from amino acid sequences. Nevertheless, it remains inferior to graph-based models, which emphasize the limitations of sequence-based methods in capturing the structural and relational information inherent in proteins. (2) General graph neural networks (GNNs) outperform sequence-based approaches on most cases, highlighting the advantage of modeling peptides as graph structures. Among these models, GIN achieves the best results, surpassing both GCN and GraphSAGE, which demonstrates the ability of GNNs to effectively capture the structural and relational dependencies within peptide graphs. (3) Recent graph-based AMP prediction methods, such as sAMPpred, MP-BERT, and PGAT, further improve performance compared with general GNNs. Their superior results can be attributed to the integration of peptide-specific structural priors and spatial constraints, enabling these models to better capture biologically meaningful geometric relationships that drive antimicrobial activity (4) \method{} outperforms all baselines across all evaluation settings and peptide length intervals, achieving the highest F1 and MCC scores in every case. This consistent superiority underscores the effectiveness of \method{}’s design, which leverages spatial structural information and mitigates class imbalance through weighted-enhanced contrastive learning and pseudo-label refinement. Specifically, the Weight-enhanced Contrastive Learning module strengthens AMP-specific structural discrimination under imbalanced conditions, while the Weight-enhanced Pseudo-label Distillation module iteratively refines pseudo labels to preserve semantic consistency and improve minority-class representation. Together, these modules enable \method{} to achieve robust, balanced, and discriminative AMP classification across diverse peptide lengths and imbalanced datasets.}

We further evaluate the flexibility of the proposed \method{} by replacing different GNN architectures as the backbone of the Graph Encoder. Specifically, GIN is replaced with GCN and GraphSAGE, and the corresponding performance results are presented in Table \ref{tab:different gnn}. The results demonstrate that \method{} w GIN consistently achieves the best performance across all metrics. This superior performance underscores GIN's strong representational capacity, enabling it to effectively capture the structural and relational characteristics of graph-structured data. This phenomenon also justifies using GIN as the backbone of Graph Encoder to enhance the proposed \method{} performance. Additionally, the visualization results among \method{} and general GNNs are shown in Figure \ref{fig:amp_nonamp_tsne}, the embeddings produced by SGAC exhibit a clear separation between AMPs and non-AMPs, indicating that the model effectively learns discriminative representations. In contrast, the embeddings from general GNNs show a less defined boundary, further highlighting the contribution of the learning modules proposed in \method{}.

\begin{table}[t] 
\centering 
\caption{The results of ablation studies. \textbf{Bold} results indicate the best performance per column.}
\resizebox{0.8\linewidth}{!}{
\begin{tabular}{lcc} 
\hline 
\textbf{Method} & \textbf{F1} & \textbf{MCC} \\ 
\hline 
\method{} w/o cf    & 74.17 $\pm$ 0.41 & 72.19 $\pm$ 0.33 \\
\method{} w/o pl    & 93.87 $\pm$ 0.47 & 91.19 $\pm$ 0.29 \\
\method{} w/o cl    & 93.74 $\pm$ 0.38 & 91.38 $\pm$ 0.31 \\
\method{} w/o all   & 92.12 $\pm$ 0.74 & 85.33 $\pm$ 0.59 \\
\method{}           & \textbf{94.83 $\pm$ 0.77} & \textbf{93.02 $\pm$ 0.56} \\
\hline
\end{tabular}
}
\label{tab:ablation_study} 

\end{table}

\subsection{3.3 Ablation study}

We conduct ablation studies to evaluate the contribution of each component in the proposed \method{}:
(1) \method{} w/o cf: removes the classification loss $\mathcal{L}_{cf}$; (2) \method{} w/o pl: removes the Weight-enhanced Pseudo-label Distillation loss $\mathcal{L}_{pl}$; (3) \method{} w/o cl: removes the Weight-enhanced Contrastive Learning loss $\mathcal{L}_{cl}$; (4) \method{} w/o all: removes all auxiliary objectives, retaining only the encoder and classifier.

Experimental results are shown in Table \ref{tab:ablation_study}. From the table, we find that 
(1) \method{} w/o cf shows lower performance compared to \method{}, showing that $\mathcal{L}_{cf}$ is crucial in leveraging labeled data effectively. This result also indicates that $\mathcal{L}_{cf}$ is crucial in guiding the model to learn meaningful and robust graph representations. (2) \method{} w/o pl exhibits a noticeable decline in performance compared to \method{}, confirming the critical role of the Weight-enhanced Pseudo-label Distillation module. This component refines pseudo labels for unlabeled or uncertain peptides according to prediction confidence and class distribution, allowing the model to leverage high-confidence samples while maintaining balanced learning between AMPs and non-AMPs. By exposing the model to a broader and more informative sample space, this mechanism improves generalization under severe class imbalance. (3) \method{} also consistently outperforms \method{} w/o cl across all metrics, demonstrating the effectiveness of the Weight-enhanced Contrastive Learning module. By assigning adaptive weights to sample pairs based on class frequency, this module enhances intra-class cohesion and inter-class separation, ensuring that minority AMP samples are adequately represented in the embedding space. This adaptive weighting effectively mitigates bias toward the dominant non-AMP class and strengthens the discriminative power of the learned representations. (4) \method{} w/o all exhibits the lowest performance, underscoring their synergistic importance. Without the Weight-enhanced Contrastive Learning and Pseudo-label Distillation components, the model fails to balance class distributions or capture subtle structural distinctions between AMPs and non-AMPs. Consequently, its learned representations become biased and less discriminative, leading to substantial drops in both F1 and MCC. 

\begin{figure}[t]
    \centering
    \includegraphics[width=0.5\textwidth]{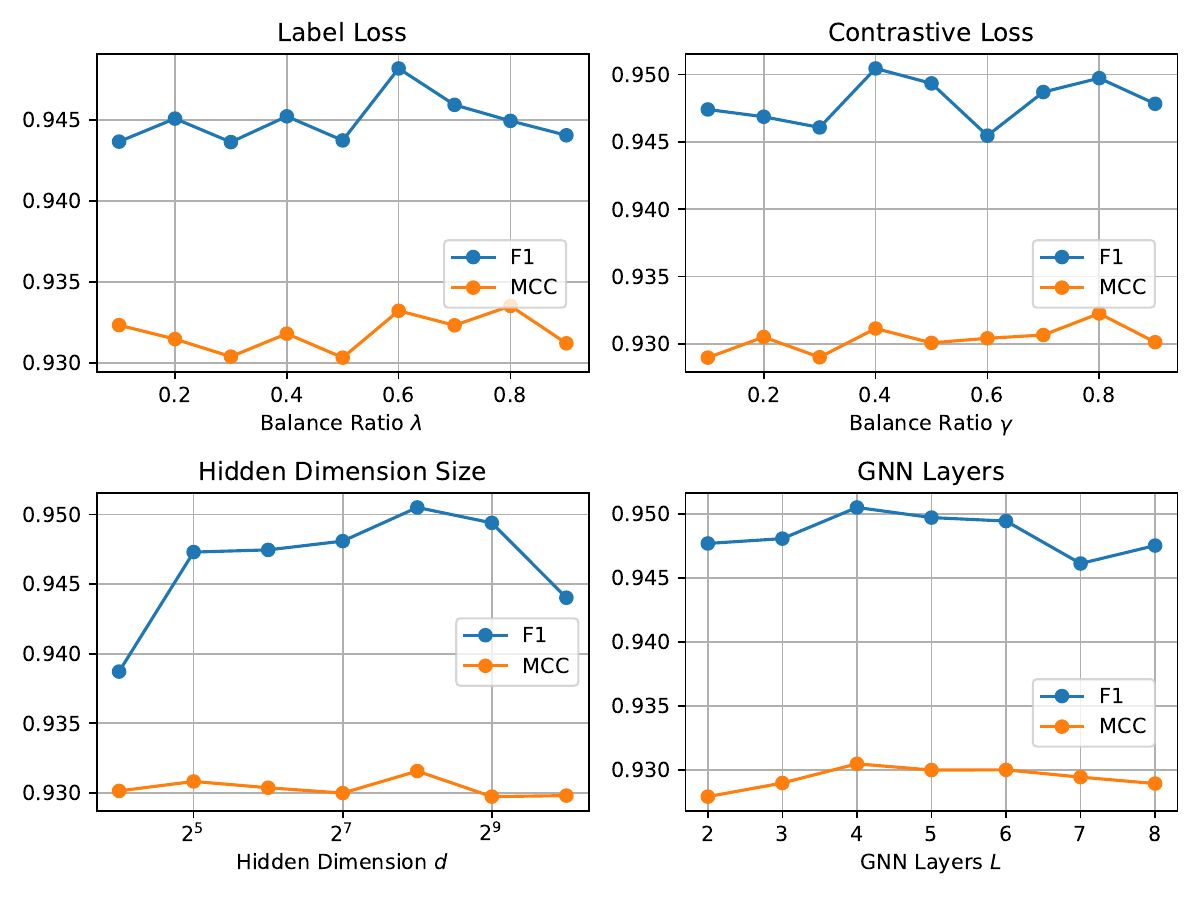} 
    \caption{Hyper-parameter sensitivity analysis of $\lambda$, $\gamma$, Hidden dimension size $d$ and GNN layers $L$.}
    \label{fig:hyper-parameters}
\end{figure}

\subsection{3.4 Sensitivity Analysis}

We study the sensitivity analysis of \method{} with respect to the impact of its hyperparameters: balance ratio $\lambda$ and $\gamma$, hidden dimension size $d$, and GNN layers $L$, which plays a crucial role in the performance of \method{}. Especially, $\lambda$ and $\gamma$ control the relative contributions of $\mathcal{L}_{\text{cl}}$ and $\mathcal{L}_{\text{pl}}$ to the overall objective, ensuring practical trade-offs between classification, contrastive learning, and pseudo-label distillation; $d$ determines the dimension of the learned graph representations, which directly impacts the model's capacity to capture structural and relational information; $L$ governs the depth of message passing in the GNNs, which influences the ability to aggregate information from neighbors and model higher-order relationships in the graph.

{Figure \ref{fig:hyper-parameters} illustrates how $\lambda$, $\gamma$, $d$ and $L$ affects the performance of \method{}. We vary $\lambda$ and $\gamma$ within the range of $\{0.1, 0.2, 0.3, 0.4, 0.5, 0.6, 0.7, 0.8,0.9\}$, $d$ in $\{2^4,2^5,2^6,2^7,2^8,2^9,2^{10}\}$, and $L$ in $\{2,3,4,5,6,7,8\}$. From the results, we observe that: (1) The performance of \method{} in Figure~\ref{fig:hyper-parameters} generally stabilizes across a wide range of $\lambda$ values and $\gamma$ values. The slight variations in the F1 score indicate a potential trade-off among the contributions of $\mathcal{L}_{cf}$, $\mathcal{L}_{cl}$ and $\mathcal{L}_{pl}$. \method{} achieves optimal performance around $\lambda=0.6$ and $\gamma = 0.4$, which is chosen as the default setting. (2) The performance of \method{} improves, as $d$ increases from $2^4$ to $2^8$, peaking at $d=2^8$. Then, the F1 score slightly declines, while AUC and PR-AUC plateau. This indicates that larger dimensions may introduce overfitting or unnecessary complexity without yielding substantial performance improvements. Therefore, $d=2^8$ is selected as the optimal hidden dimension size. (3) The performance of \method{} improves and then keeps stable as $L$ increases from 2 to 6, demonstrating that deeper architectures effectively capture structural information of protein graphs. However, when $L>6$, the model performance slightly drops due to over-smoothing. $L=4$ is selected as the default value to balance performance and computational efficiency.}

\begin{table}[t]
    \centering
    \small
    \caption{Statistics of the dataset used in the case study, showing the number of graphs, average nodes, and average edges for AMPs, non-AMPs, and their combined dataset.}
    \begin{tabular}{lcccc}
        \toprule
        Datasets       & Graphs  & Avg. Nodes & Avg. Edges & Classes \\
        \midrule
        AMPs \& non-AMPs & 78  & 23.1       & 51.7      & 2       \\
        AMPs            & 55   & 23.3       & 52.3       & 2       \\
        non-AMPs        & 23  & 22.5       & 50.1      & 2       \\
        \bottomrule
    \end{tabular}
    \label{tab:statistics_case_study}
\end{table}

\begin{table}[t]
\centering
\caption{Analysis of Misclassified Samples and Closest Structural Matches.}
\resizebox{\linewidth}{!}{%
\begin{tabular}{c|c|c|c}
\hline
\textbf{Misclassified ID} & \textbf{True $\rightarrow$ Pred} & \textbf{Closest Match ID} & \textbf{RMSD} \\
\hline
424290 & 1 $\rightarrow$ 0 & 2906279 & 1.2891 \\
425545 & 1 $\rightarrow$ 0 & 2906672 & 0.6041 \\
426383 & 0 $\rightarrow$ 1 & DRAMP04626 & 0.8319 \\
434399 & 0 $\rightarrow$ 1 & DRAMP35858 & 0.6786 \\
439751 & 0 $\rightarrow$ 1 & DRAMP33369 & 0.7193 \\
\hline
\end{tabular}
}
\label{tab:misclassified_rmsd}

\end{table}

\subsection{3.5 Case study}
To test the practicality of SGAC, we conducted a case study utilizing the potential antimicrobial peptide data synthesized in~\cite{torres2024mining}. They utilized MetaProdigal to identify 444,000 small open reading frames (smORFs) from 1,773 human microbiome metagenomes. Subsequently, they utilized AmPEP, a random forest classifier, to predict the antimicrobial activity of these smORFs, resulting in 323 smORF-encoded candidate antimicrobial peptides (SEPs). Through a series of selection criteria, they selected 78 high-priority antimicrobial peptides for synthesis and experimental validation. The vitro experiments indicate that 70.5\% (55/78) of the synthesized SEPs exhibited antimicrobial activity against at least one pathogen. Following the preprocessing methodology detailed in Section \ref{sec:3.1}, graphs are constructed for these peptides, where the statistics of these graphs are summarized in Table \ref{tab:statistics_case_study}. These peptide graphs are then input into the trained \method{} model for inference. The experimental results show that \method{} accurately classified 40 of the true AMPs and 5 of the non-AMPs.

To better understand this phenomenon, we investigate model behavior on these misclassified samples. Specifically, we randomly select five misclassified peptides and evaluate their structural similarity to samples from the training set using the Root Mean Square Deviation (RMSD) between C$_\alpha$ coordinates. We identify the closest structural match from the training data for each misclassified sample. The results are presented in Table~\ref{tab:misclassified_rmsd}, including the misclassified sample ID, the misprediction direction, the matched sample ID, and the corresponding RMSD score. These findings suggest that peptides with highly similar spatial structures (RMSD $<$ 1.5) may still exhibit distinct biological functions, leading to misclassification. Furthermore, high structural similarity can result in intrinsic representational overlap between AMP and non-AMP samples, making it difficult for the model to distinguish them with high confidence. In future work, we plan to incorporate additional biological properties, such as physicochemical characteristics, sequence motifs, and functional annotations, to enhance the robustness and interpretability of peptide classification models.

%% file: 5_conclusion.tex
\section{Conclusion}

{In this study, we proposed \method{}, a Spatial GNN-based framework for distinguishing antimicrobial peptides (AMPs) from non-AMPs by leveraging peptide three-dimensional structural information. \method{} constructs compact C$_\alpha$-based peptide graphs predicted by OmegaFold and integrates two key modules, Weight-enhanced Contrastive Learning and Weight-enhanced Pseudo-label Distillation, to address the challenges of class imbalance and limited labeled data. Through adaptive weighting and iterative pseudo-label refinement, \method{} achieves balanced, discriminative, and consistent representation learning. Extensive experiments on public AMP datasets demonstrate that \method{} outperforms both sequence-based and graph-based baselines, achieving state-of-the-art performance. Ablation and sensitivity analyses further verify the contributions of each component, and a biosynthetic case study highlights \method{}’s potential in accelerating peptide discovery and supporting antibiotic research. In future work, we plan to incorporate additional biological properties, such as physicochemical characteristics, sequence motifs, and functional annotations, to enhance the robustness and interpretability of peptide classification models.}